\documentclass[11pt]{pazh} 

\usepackage[T2A]{fontenc}
\usepackage{graphics}
\usepackage{graphicx}
\usepackage{latexsym}
\usepackage{amssymb}
\usepackage{amsmath}
\usepackage{hyperref}
\usepackage{soul}

\begin{document}

\title{\bf SDSS J170745+302056: a low surface brightness galaxy in a group}

\author{V.P. Reshetnikov\inst{1}, S.S. Savchenko\inst{1},  
A.V. Moiseev\inst{2,3,4}, O.V. Egorov\inst{2,4}}

\institute{St.Petersburg State University, Universitetskii pr. 28, St.Petersburg, 
198504 Russia
\and
Special Astrophysical Observatory, Russian Academy of Sciences, Nizhnii Arkhyz,
369167 Russia
\and
Space Research Institute, Russian Academy of Sciences, 
Profsoyuznaya 84/32, 117997 Moscow, Russia
\and
Sternberg Astronomical Institute of M.V. Lomonosov Moscow State University,
Universitetskii pr. 13, 119234 Moscow, Russia 
}

\abstract{On the basis of the SDSS survey and spectral observations
with the 6-m telescope of SAO RAS, we have peformed a detailed
study of SDSS J170745+302056. By combination of its characteristics --
exponential surface brightness distribution, central surface brightness
of stellar disk $\mu_0(B) = 23\fm25/\square''$, blue colors,
low metallicity and low star formation rate -- the galaxy is
a typical low surface brightness spiral galaxy.
Exponential scalelength of the galaxy is $\approx$3 kpc, its optical diameter
exceeds 20 kpc. SDSS J170745+302056 is a member of a group of five galaxies
and probably it is in interaction with UGC 10716. The existence of a
large low surface brightness galaxy in such a dense environment is very
unusual.
}
\titlerunning{SDSS J170745+302056}
\maketitle

\section{Introduction}

Low surface brightness (LSB) spiral galaxies are objects with central surface 
brightnesses of their stellar disks fainter than 23$^m$/$\square''$ in the $B$ 
band (Impey and Bothun 1997). This means that even the central regions of such 
galaxies are fainter than the night sky background. LSB galaxies exhibit a very wide
spread in morphology, sizes, and masses, but disk galaxies of late morphological 
types are encountered most frequently among them. LSB galaxies are rich
in gas; relatively low metallicities, blue colors, and moderate star formation 
rates are typical for them (see, e.g., van der Hulst et al. 1993; McGaugh 1994;
de Blok et al. 1995). There is some evidence that dark halos dominate in the dynamics 
of such galaxies (de Blok and McGaugh 1997).

Despite their low surface brightnesses and luminosities, the LSB galaxies are 
an appreciable component of the Universe. For example, in the local  
Universe  about a half of all galaxies with a neutral hydrogen mass 
exceeding 10$^8$ M$_{\odot}$ may belong to the LSB ones (Minchin et al. 2004). Their
contribution to the dynamical mass density produced
by all galaxies reaches $\sim$20\%, the contribution to the
density of baryonic matter is $\sim$10\%, they contribute
about a third to the total HI density in the nearby
Universe (Minchin et al. 2004).

The origin and evolution of LSB galaxies remain largely unclear. A number of 
causes were suggested to explain their peculiarities (primarily a large amount
of gas and blue colors in combination with a low star formation efficiency), 
such as, for example, the formation of LSB galaxies inside relatively late-formed 
dark halos with a large spin (Jimenez et al. 1998) or a low
metallicity of the interstellar medium preventing an efficient cooling and creating 
a deficit of cold molecular gas  which is fuel  for star formation (Gerritsen and
de Blok 1999). In recent years, the most popular model
has been the scenario of small short starbursts randomly distributed over the 
galactic disk (de Blok et al. 1995; Gerritsen and de Blok 1999; Vorobyov et al.
2009). The mechanism controlling the sporadic star formation in LSB galaxies is 
unknown, but this model is capable of reproducing many of their characteristics,
including their blue colors and low metallicities (Vorobyov et al. 2009).

Our paper is devoted to an observational study of the LSB galaxy SDSS J170745+302056. 
In the second section of the paper we present the results of
its detailed study based on SDSS data and our own spectroscopic observations, 
and in the third section we consider its spatial environment. All magnitudes
in the paper are given in the AB system.

\section{General characteristics of SDSS J170745+302056}

\subsection{Observations and spectroscopic parameters}

SDSS J170745+302056 was discovered serendipitously during spectroscopic 
observations of the galaxy UGC 10716. The observations of UGC 10716
were performed in July 2013 as part of our program aimed at studying edge-on 
spiral galaxies with strongly warped stellar disks (Reshetnikov et
al. 2016). A spectrum of the galaxy in the red region (5700 -- 7500 \AA) 
with the slit position angle P.A. = 12$^\circ$ was taken with the SCORPIO
instrument (Afanasiev and Moiseev 2005) at the 6-m SAO RAS telescope (the 
parameters of these observations and the data reduction are described in 
Reshetnikov et al. 2016). Apart from UGC 10716, the spectrograph
slit crossed a faint galaxy about 80$\arcsec$ north of it. This
galaxy, SDSS J170745+302056, is barely seen on the original SDSS frames but 
becomes prominent when increasing the contrast (Fig. 1).

\begin{figure}
\centering
\includegraphics[width=0.5\textwidth, angle=0, clip=]{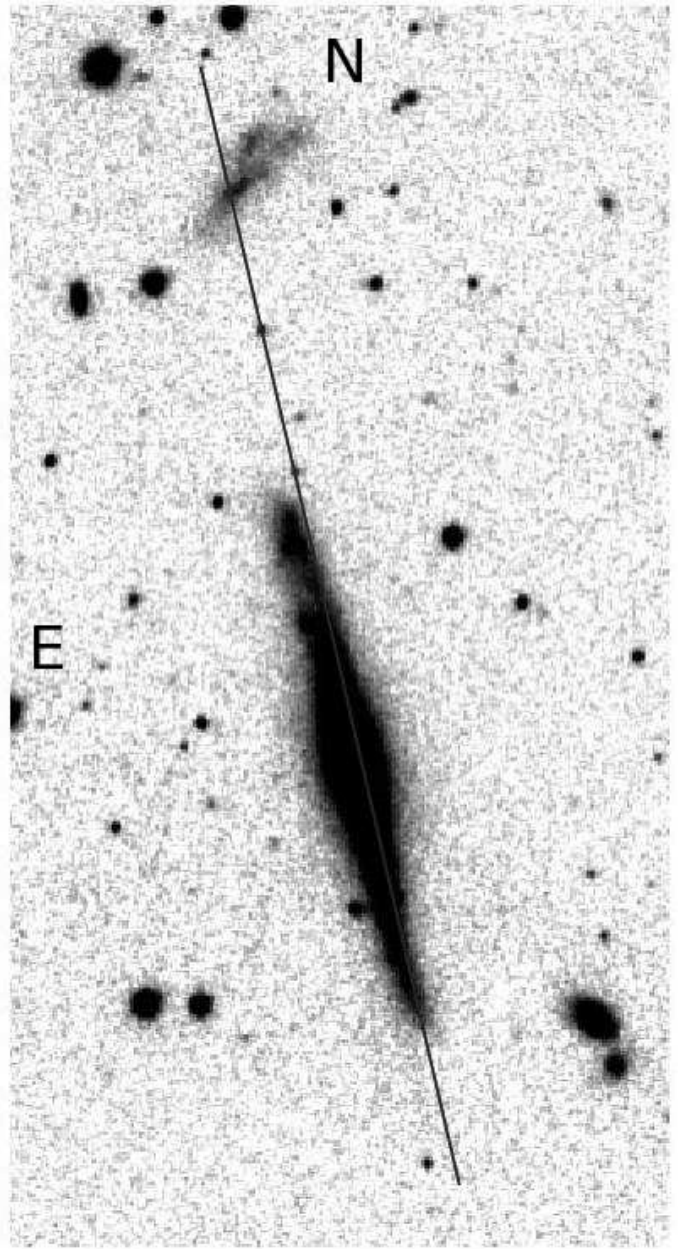}\\
\caption{The high-contrast combined ($g + r + i$) image of 
UGC 10716 (large edge-on galaxy) and SDSS J170745+302056 (small galaxy at the top). 
The straight line indicates the spectrograph slit position. The frame is 1.\arcmin55 $\times$ 2.\arcmin93 in size.
}
\end{figure}

The spectrograph slit passed almost through the nucleus of SDSS J170745+302056, 
approximately at 1$\arcsec$ from the bright knot near the image center. The
H$\alpha$ emission line, from which we found the radial velocity of the galaxy 
to be V$_{sys}$ = 9453$\pm$5 km/s, and noticeably weaker [NII], [SII], and HeI 
lines are clearly seen in its spectrum (see the integrated spectrum
of the central part of the galaxy in Fig. 2). This velocity is lower than the 
radial velocity of the nucleus of UGC 10716 by 232 km/s. Given that
SDSS J170745+302056 and UGC 10716 are close in projection, it can be concluded 
that they form a bound system or, as will be shown below, enter into
a common group of galaxies.

\begin{figure*}
\centering
\includegraphics[width=0.9\textwidth, angle=0, clip=]{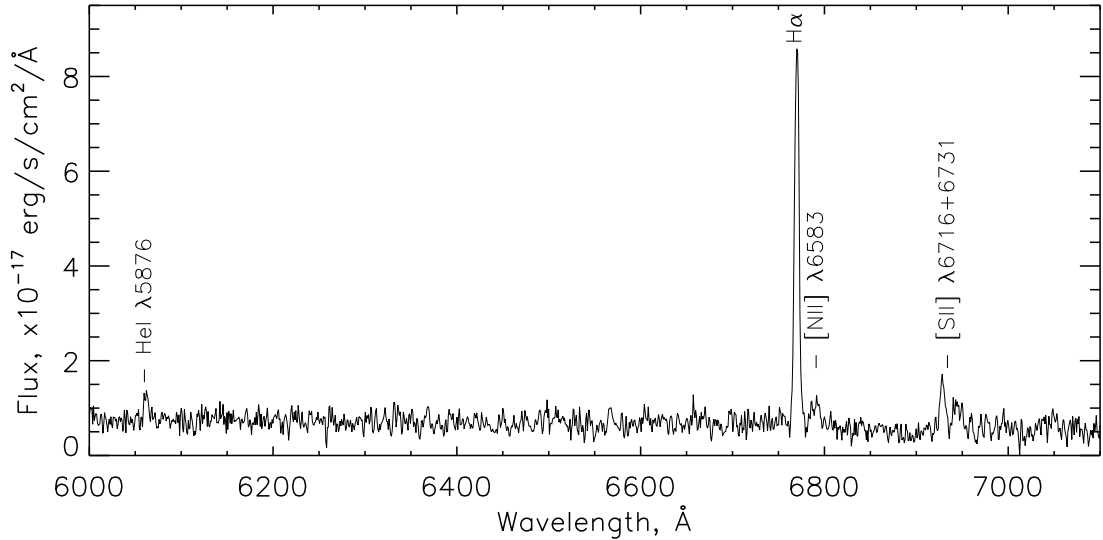}\\
\caption{Observed spectrum of the central region of SDSS J170745+302056.}
\end{figure*}

The observed line fluxes from the integrated spectrum of the galaxy are 
listed in Table 1. Using the calibration from Marino et al. (2013), from the
ratios of the [NII]$\lambda$6583 \AA\,\, and H$\alpha$ intensities we
estimated the gas metallicity in the galaxy to be 12+log(O/H) = 8.20$\pm$0.05 
or $\approx$1/3 of the solar value. It should be noted, however, that we estimated 
the above error in the oxygen abundance based on the formal errors in the line 
fluxes from Table 1. The error of our metallicity estimation method reaches a factor
of $\sim$2, so that the actual error can be higher by several times.

\begin{table}
\caption{Line fluxes in the spectrum of SDSS J170745+302056} 
\begin{center}
\begin{tabular}{|ll|}
\hline 
Line & Flux, 10$^{-16}$ erg/s/cm$^2)$ \\
\hline
HeI$\lambda$5876     & 0.39 $\pm$ 0.08     \\
$[$NII$]\lambda$6548 & 0.12 $\pm$ 0.08      \\
H$\alpha$            & 5.5  $\pm$ 0.2      \\
$[$NII$]\lambda$6584 & 0.37 $\pm$ 0.07      \\
$[$SII$]\lambda$6717 & 0.83 $\pm$ 0.09      \\
$[$SII$]\lambda$6731 & 0.47 $\pm$ 0.08      \\
\hline
\end{tabular}
\end{center}
\end{table}

The adopted photometric distance to the galaxy is 133 Mpc (see Section 3 of 
this paper), giving a linear scale of 0.61 kpc/$\arcsec$.

\subsection{Photometric characteristics}

The basic observed characteristics of SDSS J170745+302056 are summarized in Table 2.
The integrated apparent magnitudes of the galaxy were derived from the original 
SDSS frames\footnote{http://www.sdss.org/dr12/} by integrating the flux within 
an elliptical aperture with a semimajor axis of 22\arcsec. (At the preliminary 
stage we subtracted the background found by a large region
around the galaxy and masked the nearby stars.) On the whole, these magnitudes 
are close to those in SDSS: the mean difference between the SDSS data
in five filters and our measurements is $+0\fm05 \pm 0\fm18$.

\begin{table}[h]
\caption{Observed characteristics of SDSS J170745+302056} 
\begin{center}
\begin{tabular}{| l |  c  | }
\hline
$\alpha$ (2000)  & 17:07:45.80 \\ 
$\delta$ (2000)  & +30:20:56.2 \\
\hline
$u$   &  19.63 $\pm$ 0.16   \\
$g$   &  18.72 $\pm$ 0.06   \\
$r$   &  18.46 $\pm$ 0.06   \\
$i$   &  18.38 $\pm$ 0.10   \\
$z$   &  18.52 $\pm$ 0.36   \\
D (Mpc) & 133  \\
Scale & 0.61 kpc/1$\arcsec$     \\
$M_B$ &  --16.9    \\
d($\mu(g)=26.5$) & 36$\arcsec$ (22 kpc)  \\
$\mu_0 (g)$ & 23.28 $\pm$ 0.05  \\
$\mu_0 (r)$ & 22.91 $\pm$ 0.10  \\
$\mu_0 (i)$ & 22.86 $\pm$ 0.06  \\
$h(g)$      &  6.$\arcsec$4 $\pm$ 0.$\arcsec$2  \\
$h(r)$      &  4.$\arcsec$9 $\pm$ 0.$\arcsec$2     \\
$h(i)$      &  5.$\arcsec$5 $\pm$ 0.$\arcsec$2    \\
12 + lg(O/H) & 8.20 $\pm$ 0.05 \\
\hline
\end{tabular}
\end{center}
\end{table}

SDSS J170745+302056 has an irregular morphology (Figs. 1 and 3) and, therefore, 
we used averaging along ellipses to construct its photometric profiles.
The positions of the centers of the ellipses and the position angles of their 
major axes were identical in all color bands. The cuts in the $g$, $r$, and $i$ 
filters constructed in this way are shown in Fig. 3. The semimajor axis of the 
ellipse and the corresponding mean surface brightness are plotted along the horizontal
and vertical axes of the figure, respectively. Since the galaxy is blue, the cuts in 
different bands overlap, and, therefore, for convenience, the data in the $r$ 
and $i$ filters in Fig. 3 were shifted toward higher brightnesses.

To a first approximation, the averaged brightness distribution of SDSS J170745+302056 
is described by an exponential law with a scale length
$h = 5\farcs6 \pm 0\farcs75$ = 3.4 kpc (the value averaged over three filters).
The galaxy$'$s surface brightness in the $u$ and $z$ bands fluctuates strongly, 
but, on the whole, it can also be fitted by an exponential disk model with a scale 
length of $6\farcs8 \pm 0\farcs5$ ($u$) or $5\farcs6 \pm 0\farcs5$ ($z$). 
A slight brightness excess that is most likely related to the extended
structure north of the galactic center is noticeable in
the photometric profiles at a distance of $\sim$10$\arcsec$--15$\arcsec$
(Fig. 1). If this feature in the profiles is eliminated,
the exponential disk scale length decreases to 
$h = 5\farcs1 \pm 0\farcs8$ = 3.1 kpc (the value averaged over the
$g$, $r$, and $i$ filters), while the central surface brightness
remains virtually without changes. Table 2 gives the values of $h$ and $\mu_0$ 
found from the brightness profiles without eliminating the excess region.

\begin{figure}
\centering
\includegraphics[width=0.38\textwidth, angle=-90, clip=]{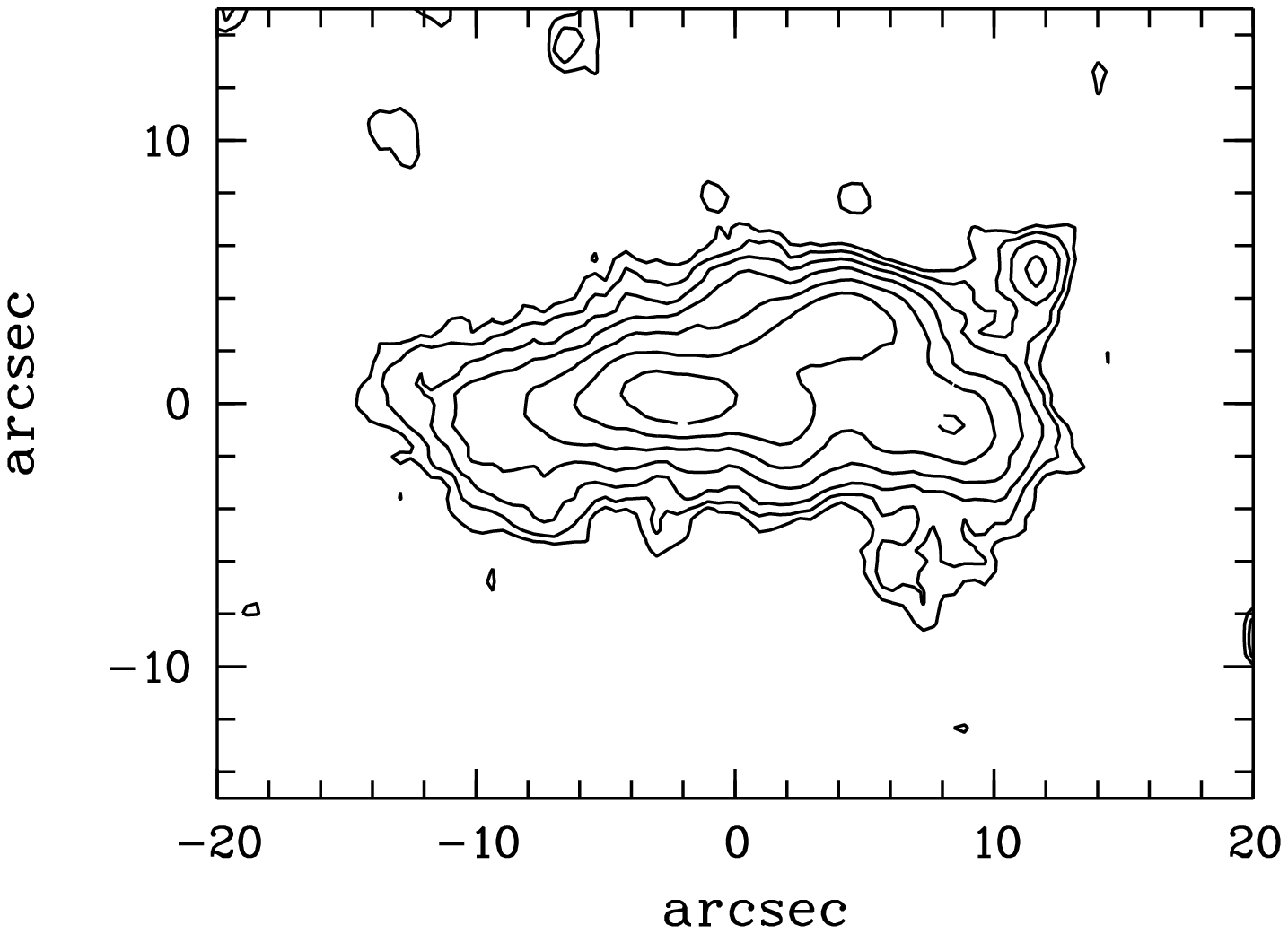}\\
\includegraphics[width=0.38\textwidth, angle=-90, clip=]{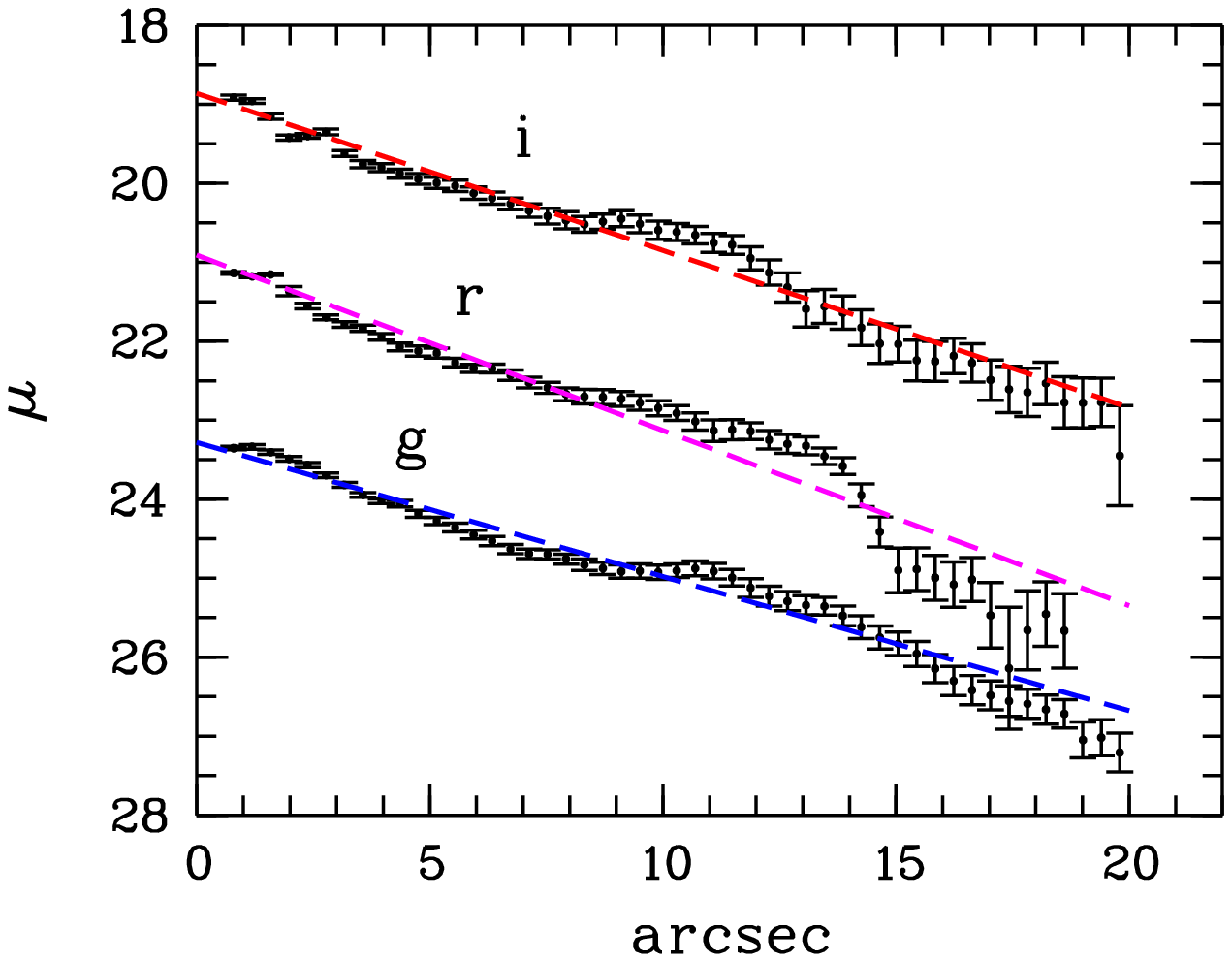}\\
\caption{{\it Top}: isophotal map of SDSS J170745+302056 constructed from the 
combined ($g$ + $r$ + $i$) image of the galaxy. The galaxy image was rotated so that 
its major axis coincids with the horizontal axis of the figure. 
{\it Bottom}:  elliptically 
averaged photometric profiles along the major axis of SDSS J170745+302056. 
The dashed lines indicate the fits to the cuts by an exponential disk model. 
The surface brightnesses in the $r$ and $i$ filters were shifted relative to 
the $g$ filter by 2$^m/\square''$ and 4$^m/\square''$, respectively.}
\end{figure}

The observed central surface brightness of the SDSS J170745+302056 disk (Table 2) 
after its correction for extinction in the Milky Way (Schlafly and
Finkbeiner 2011), applying the k-correction (Chilingarian et al. 2010), and the 
recalculation to the $B$ filter (Blanton and Roweis 2007) is
$\mu_0(B) = 23.25$$^m$/$\square''$. Consequently, the galaxy is indeed a LSB one. 
The galaxy is not bright, its absolute $B$ magnitude is --16.9, but it cannot 
be attributed to dwarf objects by its diameter $\sim$20 kpc.

The metallicity of SDSS J170745+302056 corresponds to typical values for LSB 
objects (e.g. McGaugh 1994; Kuzio de Naray et al. 2004). The characteristics of 
SDSS J170745+302056 satisfy the luminosity -- metallicity relation for spiral
galaxies (Kuzio de Naray et al. 2004).

\subsection{Analysis of the colors}

Figure 4 shows the positions of the corrected colors for SDSS J170745+302056 on 
color--color diagrams. On the right panel the observations in the
$NUV$ filter ($\lambda$ = 0.23 $\mu$m) based on data from the
GALEX space telescope (Martin et al. 2005) were added to the optical photometry. 
The dashed lines on the panels indicate the dependences for the colors
of normal galaxies. On the left panel we used the SDSS data for bright E--Im type 
galaxies (Fukugita et al. 2007). On the right panel we added the
$NUV-r$ color index according to the empirical dependence from Chilingarian and 
Zolotukhin (2012) to the $g-r$ color index from Fukugita et al. (2007)
for galaxies of the same morphological types. We see that the integrated colors 
of SDSS J170745+302056 are very blue, $(g-r)_0 = +0.26 \pm 0.08$, 
$(u-g)_0 = +0.72 \pm 0.17$, and $(NUV-r)_0 = +1.61 \pm 0.17$, and
are outside the color sequences for the Hubble sequence of galaxies.

\begin{figure*}
\centering
\includegraphics[width=0.45\textwidth, angle=-90, clip=]{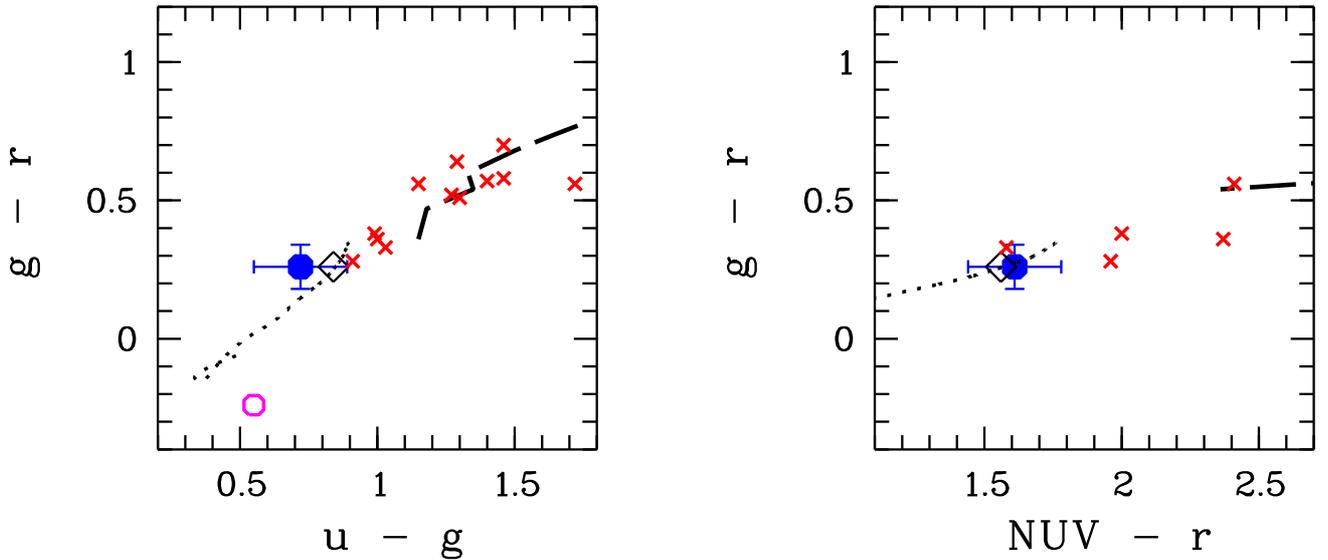}\\
\caption{{\it Left}: the colors of SDSS J170745+302056 on the $(g-r)$ -- $(u-g)$ 
color--color diagram (blue circle with error bars). The dashed line is the color 
sequence for normal galaxies (Fukugita et al. 2007). The dotted line is the 
evolutionary track for the model with constant star formation (see the text). 
The diamonds indicates the colors of the model for an age of 6.5 Gyr. The circle
marks the position of the galaxy UGC 12695. {\it Right}: the colors of 
SDSS J170745+302056 on the $(g-r)$ -- $(NUV-r)$ color--color diagram. The dashed 
line indicates the colors of normal galaxies (Fukugita et al. 2007; 
Chilingarian and Zolotukhin 2012). The remaining designations are the same as 
those on the left panel. The red crosses on both panels indicate the characteristics of
LSB galaxies from Du et al. (2017).}
\end{figure*}

The crosses in Fig. 4 indicate the characteristics of edge-on LSB galaxies from 
Du et al. (2017). These galaxies have metallicities, luminosities, and sizes
comparable to those of SDSS J170745+302056. As can be seen from Fig. 4, the LSB 
galaxies are located on the color--color diagrams approximately along
the color sequence for normal galaxies, but with a shift toward bluer colors. 
On these diagrams SDSS J170745+302056 lies at the edge of the distributions
for LSB galaxies. The circle on the $(g-r)$ -- $(u-g)$ plane indicates the 
colors of UGC 12695, one of the bluest known LSB galaxies (O$'$Neil et al. 2000).
Figure 4 suggests that SDSS J170745+302056 belongs to very, but not extremely 
blue LSB galaxies.

To interpret the observed colors of SDSS J170745+302056 using the GALEV software
package\footnote{http://www.galev.org} (Kotulla et al. 2009), we considered a
simple evolutionary model. This model is based on the assumption about a constant 
star formation rate over the entire lifetime of the galaxy. This assumption,
of course, greatly simplifies the actual star formation history, but for many 
LSB galaxies it can hold, even if approximately (Schombert and McGaugh 2014).
Applying a more complex and refined modeling to a scarce observational data set 
and relatively high measurement errors was deemed premature.

To estimate the star formation rate in SDSS J170745+302056, we used its ultraviolet
luminosity in the $FUV$ ($\lambda=0.15$ $\mu$m) and $NUV$ filters from the GALEX 
data. (The luminosity in this range traces the star formation on a time scale
of $\sim10^8$ years.) We recalculated the luminosities to the
star formation rate based on the calibrations from Wijesinghe et al. (2011). 
The star formation rate was found from the data in the $FUV$ and $NUV$ filters
to be 0.04 and 0.05 M$_{\sun}$/yr, respectively.

This estimate is consistent with the estimates based on the observed flux in the 
H$\alpha$ emission line in our spectroscopic observations (Table 1). Using the
calibration from Kennicutt (1998), given the extinction in the Milky Way and 
the adopted distance, we obtain the current star formation rate, 0.01 M$_{\sun}$/yr.
Of course, this value is a lower limit, for example, other star-forming regions 
through which the spectrograph slit did not pass can also be present in the galaxy.

Assuming the star formation rate in SDSS J170745+302056 to be constant and equal to
0.05 M$_{\sun}$/yr, we computed the long-term evolution of its colors. Other 
components of the model were: the Salpeter initial mass function, 
emission lines contribution, the absence of internal extinction in the galaxy, and consistent 
evolution of the chemical composition (Kotulla 2009). The final tracks on the
color--color diagrams are indicated by the dotted lines (Fig. 4).

On both diagrams the model tracks pass near the colors of SDSS J170745+302056. 
The best agreement between the model and actual colors is achieved
for a time of 6.5 Gyr after the onset of star formation in the galaxy (the 
diamonds in Fig. 4). By this time, the stellar mass
M$_* = 3.25 \cdot 10^8$ M$_{\odot}$ should be accumulated in the galaxy and 
its metallicity should be $\sim$1/10 of the solar one. The actually observed
metallicity of SDSS J170745+302056 is approximately a factor of 3 higher than 
this value. However, given the error in determining the metallicity in the
galaxy and that the model used is approximate, the question about the agreement 
between the model and observations remains open. The stellar mass of
SDSS J170745+302056 can be estimated from its multicolor photometry. Using the 
calibrations from Bell et al. (2003) and Zibetti et al. (2009) and various
colors, we found the stellar mass of the galaxy to reach $(2-6) \cdot 10^8$ M$_{\odot}$. 
This value is consistent with  the model above.

\section{The spatial environment of SDSS J170745+302056}

SDSS J170745+302056 is not far from the large and bright spiral galaxy 
UGC 10716 (Fig. 1). The projected distance between the centers of the
galaxies is 84$\arcsec$ or 51 kpc. As we have shown previously, the radial 
velocities of the galaxies are relatively close, which may be indicative of their
interaction. Additional arguments for the possibility of an interaction between 
these galaxies can be the stellar disk warp in UGC 10716 (Reshetnikov
et al. 2016) and the peculiar morphology of SDSS J170745+302056. Furthermore, 
the combined image of SDSS J170745+302056 shows signatures
of a small tidal tail branching off from the northern edge of the galaxy toward 
UGC 10716 (Fig. 3).

According to Berlind et al. (2006), UGC 10716 is a member of a group of three 
(from a sample of objects with absolute magnitudes $M_r < -18$, group 5002)
or four (from a sample with $M_r < -19$, group 9524) galaxies. Apart from 
UGC 10716, the group includes PGC 059633 (the brightest member of the group),
UGC 10714, and SDSS J170726.32+301316.7 (Berlind et al. 2006). A check based on SDSS
data shows that SDSS J170726.32+301316.7 has no measured redshift and is most 
likely is a more distant background object that does not belong
to this group. On the other hand, the irregular galaxy SDSS J170730.04+301356.8 
not far from UGC 10714 and the galaxy SDSS J170745+302056
being studied in this paper should be included in the group. The corrected 
and expanded list of group members is presented in Table 3. The data in the 
table were collected from SDSS and NED\footnote{http://ned.ipac.caltech.edu}; 
our results were also used.

\begin{table*}
\caption{Parameters of the group galaxies} 
\begin{center}
\begin{tabular}{|llllll|}
\hline
No. & SDSS & Name & Type & cz (km/s)  & $r$ (mag) \\ 
\hline
1 & J170654.71+301611.1 &  PGC 059633 & Sb & 9342 & 13.73  \\
2 & J170725.22+301330.0 &  UGC 10714  & Sb & 9526 & 14.28  \\
3 & J170730.04+301356.8 &             & Irr& 9503 & 17.04  \\
4 & J170744.50+301935.1 &  UGC 10716  & Sb & 9685 & 14.74  \\
5 & J170745.80+302056.2 &             & LSB, Irr   & 9453 & 18.46  \\
\hline
\end{tabular}
\end{center}
\end{table*}

The mean radial velocity of the group galaxies is 9502 km/s and the corresponding 
distance to it is D = 133 Mpc (this value was used previously to find
all of the distance-dependent quantities). The velocity dispersion of the group 
galaxies is $\sigma$ = 121 km/s; its harmonic radius is 151 kpc. The above values
are quite typical for the groups identified by various algorithms in the 
surrounding part of the Universe (Makarov and Karachentsev 2011).

The group crossing time is $\approx$0.1 of the Hubble time. This means that the 
group members could repeatedly encounter in the past. It remains unclear
how a relatively unevolved and large ($\sim$20 kpc in diameter!) LSB galaxy 
could be preserved in such an environment. LSB galaxies are usually located
in low-density enviroment, far from possible star formation triggers 
(see Rosenbaum et al. 2009 and references therein). Therefore, we can
suggest that SDSS J170745+302056 has encountered this group only relatively 
recently and, possibly, experiences the first encounter with another massive
galaxy (UGC 10716). In that case, the large-scale warp of the UGC 10716 plane 
is most likely related to the interaction with other group members and
not with SDSS J170745+302056. Remarkably, excluding SDSS J170745+302056 from 
consideration barely changes the integrated characteristics of the
group. Without SDSS J170745+302056 the velocity dispersion of the group becomes 
136 km/s, its harmonic radius is 158 kpc, and the crossing time
remains the same.

Recently, a large population of the so-called ultra-diffuse
galaxies was discovered in several clusters (van Dokkum et al. 2015; 
Wittmann et al. 2017) and even groups of galaxies (Makarov et al. 2015).
These galaxies have very low surface brightnesses, low metallicities, and their 
typical sizes reach several kiloparsecs. The origin of such objects in the regions
of a high density of galaxies also remains unclear. For example, it is possible 
that the ultra-diffuse galaxies have unusually massive (for their luminosity)
dark halos that protect them from the tidal effect of clusters 
(van Dokkum et al. 2015). Drawing an analogy between ultra-diffuse galaxies and
SDSS J170745+302056, we can suggest that the galaxy being studied also possesses 
a very massive halo that prevents its destruction. On the other
hand, compared to SDSS J170745+302056, the ultra-diffuse galaxies are fainter, 
have, on average, lower surface brightnesses, and, in addition, they
contain no gas and exhibit red colors. Consequently, SDSS J170745+302056 and 
the ultra-diffuse galaxies can be different in origin.

\section{Conclusions}

Based on our observational data, we studied the galaxy SDSS J170745+302056. 
By the set of its characteristics, it is a typical blue irregular LSB
galaxy. We estimated its age within a simple model with a constant star 
formation to be $\sim$6.5 Gyr. This means that it was formed at a 
redshift $z \sim 0.7$.

The most remarkable feature of SDSS J170745+302056 is that it is a member
of a group of galaxies with a crossing time $\approx$0.1 of the Hubble time. 
In such a dense environment the galaxy should have repeatedly encountered
other group members and experienced tidal perturbations capable of triggering 
active star formation in it. At the same time, the low surface
brightness of SDSS J170745+302056 and its colors may be indicative of the 
galaxy$'$s relatively quiet evolution in the past. It is not inconceivable that
SDSS J170745+302056 was formed in a region of relatively low ambient density 
and has only recently encountered the group of galaxies we discuss.
Another possible explanation proposed previously for ultra-diffuse galaxies 
in clusters is that SDSS J170745+302056 possesses an anomalously
massive dark halo that protects its disk during its
encounters with other group members.

Our results are preliminary ones. More detailed observational data 
concerning the morphology, kinematics, and HI content of
SDSS J170745+302056 are required to clarify its formation and evolution.

\section{Acknowledgments}

We are grateful to the referee for the useful and constructive remarks.

We used the observational data from the 6-m SAO RAS telescope
financially supported by the Ministry of Education and Science of the Russian 
Federation (agreement no. 14.619.21.0004, project IDPRFMEFI61914X0004). 

We used the NASA/IPAC Extragalactic Database (NED), which is operated by
the Jet Propulsion Laboratory, the California Institute of Technology, under 
contract with the National Aeronautics and Space Administration (USA), and
the publicly accessible SDSS data.

SDSS-III is financed by the Alfred P. Sloan Foundation, the participating 
organizations of the SDSS collaboration, the National Science Foundation, and
the US Department of Energy. SDSS-III is performed by the Astrophysical Research 
Consortium for the Participating Institutions of the SDSS-III
Collaboration, including the University of Arizona, the Brazilian Participation 
Group, the Brookhaven National Laboratory, the Carnegie Mellon University,
the University of Florida, the French Participation Group, the German 
Participation Group, the Harvard University, the Instituto de Astrofısica 
de Canarias, the Michigan State/Notre Dame/JINA Participation Group, the John 
Hopkins University, the Lawrence Berkeley National Laboratory, the Max Planck 
Institute for Astrophysics, the Max Planck Institute for Extraterrestrial 
Physics, the New Mexico State University, New York University, Ohio State University,
Pennsylvania State University, the University of Portsmouth, the Princeton 
University, the Spanish Participation Group, the University of Tokyo, the
University of Utah, the Vanderbilt University, the University of Virginia, 
the University of Washington, and the Yale University.

\pagebreak

\begin{center}
{\Large \bf References}
\end{center}

\noindent
V.L. Afanasiev, A.V. Moiseev, Astron. Lett. {\bf 31}, 194 (2005). \\

\noindent
E.F. Bell, D.H. McIntosh, N. Katz, M.D. Weinberg, 
Astrophys. J. Suppl. {\bf 149}, 289 (2003). \\

\noindent
A.A. Berlind, J. Friedman, D.H. Weinberg, et al., Astrophys. J. Suppl. {\bf 167}, 
1 (2006). \\

\noindent
M.R. Blanton, S. Roweis, Astron. J. {\bf 133}, 734 (2007).\\

\noindent
I. Chilingarian, A.-L. Melchior, I. Zolotukhin, 
Mon. Not. R. Astron. Soc. {\bf 405}, 1409 (2010). \\

\noindent
I.V. Chilingarian, I.Yu. Zolotukhin, 
Mon. Not. R. Astron. Soc. {\bf 419}, 1727 (2012). \\

\noindent
W.J.G. de Blok, J.M. van der Hulst, G.D. Bothun, 
Mon. Not. R. Astron. Soc. {\bf 274}, 235 (1995). \\

\noindent
W.J.G. de Blok, S.S. McGaugh, Mon. Not. R. Astron. Soc. {\bf 290}, 533 (1997). \\

\noindent
W. Du, H. Wu, Y. Zhu, et al., Astrophys. J. {\bf 837}, id. 152 (2017). \\

\noindent
M. Fukugita, O. Nakamura, S. Okamura, et al., Astron. J. {\bf 134}, 579 (2007). \\

\noindent
J.P.E. Gerritsen, W.J.G. de Blok, Astron. Astrophys. {\bf 342}, 655 (1999). \\

\noindent
Ch.Impey, G. Bothun, Ann. Rev. Astron. Astrophys. {\bf 35}, 267 (1997). \\

\noindent
R. Jimenez, P. Padoan, F. Matteucci, A.F. Heavens, 
Mon. Not. R. Astron. Soc. {\bf 299}, 123 (1998). \\

\noindent
R.C. Kennicutt, Ann. Rev. Astron. Astrophys. {\bf 36}, 189 (1998). \\

\noindent
R. Kotulla, U. Fritze, P. Weilbacher, P. Anders, 
Mon. Not. R. Astron. Soc. {\bf 396}, 462 (2009). \\

\noindent
R. Kuzio de Naray, S.S. McGaugh, W.J.G. de Blok, 
Mon. Not. R. Astron. Soc. {\bf 355}, 887 (2004). \\

\noindent
D. Makarov, I. Karachentsev, Mon. Not. R. Astron. Soc. {\bf 412}, 2498 (2011). \\

\noindent
D.I. Makarov, M.E. Sharina, V.E. Karachentseva, 
I.D. Karachentesev, Astron. Astrophys. {\bf 581}, A82 (2015). \\

\noindent
R.A. Marino, F.F. Rosales-Ortega, S.F. Sanchez, et al., 
Astron. Astrophys. {\bf 559}, A114 (2013). \\

\noindent
D.Ch. Martin, J. Fanson, D. Schiminovich, et al., Astrophys. J. {\bf 619}, L1 (2005). \\

\noindent
S.S. McGaugh, Astrophys. J. {\bf 426}, 135 (1994). \\

\noindent
R.F. Minchin, M.J. Disney, Q.A. Parker, et al., 
Mon. Not. R. Astron. Soc. {\bf 355}, 1303 (2004). \\

\noindent
K. O'Neil, M.A.W. Verheijen, S.S. McGaugh, Astron. J. {\bf 119}, 2154 (2000). \\

\noindent
V.P. Reshetnikov, A.V. Mosenkov, A.V. Moiseev, et al., 
Mon. Not. R. Astron. Soc. {\bf 461}, 4233 (2016). \\

\noindent
S.D. Rosenbaum, E. Krusch, D.J. Bomans, R.-J. Dettmar,
Astron. Astrophys. {\bf 504}, 807 (2009). \\

\noindent
E.F. Schlafly, D.P. Finkbeiner, Astrophys. J. {\bf 737}, 103 (2011). \\

\noindent
J. Schombert, S. McGaugh, Publ. Astron. Soc. Austr. {\bf 31}, 36 (2014).\\

\noindent
J.M. van der Hulst, E.D. Skillman, T.R. Smith, et al., 
Astron. J. {\bf 106}, 548 (1993). \\

\noindent
P.G. van Dokkum, A.J. Romanowsky, R. Abraham, et al.,
Astrophys. J. {\bf 804}, L26 (2015). \\

\noindent
E.I. Vorobyov, Yu. Shchekinov, D. Byzyaev, et al., 
Astron. Astrophys. {\bf 505}, 483 (2009). \\

\noindent
D.B. Wijesinghe, A.M. Hopkins, R. Sharp, et al., 
Mon. Not. R. Astron. Soc. {\bf 410}, 2291 (2011). \\

\noindent
C. Wittmann, Th. Lisker, L. Ambachew Tilahun, et al.,
Mon. Not. R. Astron. Soc. {\bf 470}, 1512 (1997). \\

\noindent
S. Zibetti, S. Charlot, H.-W. Rix, Mon. Not. R. Astron. Soc. {\bf 400}, 1181 (2009). \\

\end{document}